\documentclass[aps,pre,twocolumn,floatfix,nofootinbib,amsmath,amssymb]{revtex4}
\usepackage{times,epsfig}
\usepackage[sort&compress]{natbib}
\usepackage{graphicx}
\usepackage[parfill]{parskip}
\usepackage{hyperref}

\begin{document}

\title{Stochastic Total Quasi-Steady-State Approximation for the Michaelis-Menten Scheme}

\author{Vahe Galstyan$^{1}$}
\email{vg2321@columbia.edu} \affiliation{$^1$Columbia College, Columbia University, New York, NY 10027}

\begin{abstract}
In biochemical systems the Michaelis-Menten (MM) scheme is one of the best-known models of the enzyme-catalyzed kinetics. In the academic literature the MM approximation has been thoroughly studied in the context of differential equation models. At the level of the cell, however, molecular fluctuations have many important consequences, and thus, a stochastic investigation of the MM scheme is often necessary.  In their work Barik \textit{et al.} [Biophysical Journal, \textbf{95}, 3563-3574, (2008)] presented a stochastic approximation of the MM scheme. They suggested a substitution of the propensity function in the reduced master equation with the total quasi-steady-state approximation (tQSSA) rate. The justification of the substitution, however, was provided for a special case only and did not cover the whole parameter domain of the tQSSA. In this manuscript we present a derivation of the stochastic tQSSA that is valid for the entire tQSSA parameter domain.

\end{abstract}

\maketitle

\section{Introduction}

Enzyme-catalized reactions are frequently encountered in biology and are modeled with the MM scheme:

\begin{eqnarray}
\label{eqn: MMK}
E + S \underset{k_{-1}}{\overset{k_1}{\rightleftharpoons}} C {\overset{k_2}{\rightarrow}} E + P. \label{MMK}
\end{eqnarray}

Here $k_1$ and $k_{-1}$ are reaction constants for the complex ($C$) formation and dissociation respectively, and $k_2$ is the catalysis rate constant. Rate equations for the three elementary reactions in the scheme are presented through the following ODEs:

\begin{eqnarray}
\frac{dS}{dt} &=& -k_1S(E_0-C) + k_{-1}C, \label{eqn: ODE Substrate} \\
\frac{dC}{dt} &=& -(k_{-1}+k_2)C + k_1S(E_0-C) \label{eqn: ODE Complex}.
\end{eqnarray}

Here $E_0$ is the initial concentration of enzyme molecules, and the law of mass action is used to write enzyme $E$ as $(E_0-C)$, assuming $(E_0, S_0, 0, 0)$ as the inital condition.

In \cite{seg} Segel derives the deterministic quasi-steady-state approximation (QSSA) with the validity condition 
\begin{eqnarray}
\label{eqn: QSSA domain}
E_0 \ll S_0 + K_M, 
\end{eqnarray}
where $K_M = (k_{-1}+k_2)/k_1$ is the MM constant. The approximate rate law reads:

\begin{eqnarray}
\label{eqn: QSSA}
\frac{dS}{dt} = -\frac{k_2 E_0 S}{K_M + S}.
\end{eqnarray}

Tzafriri \cite{tzaf} extends the results obtained by Segel \cite{seg} by suggesting the method called the total QSSA (tQSSA). It applies to reactions where the concentration of the total substrate ($\bar{S} =S + C$) changes on a slower timescale than the concentration of the enzyme-substrate complex ($C$). TQSSA implies:

\begin{eqnarray}
\frac{d\bar{S}}{dt} &=& -k_2 C (\bar{S}), \label{eqn: TQSSA}\\
C(\bar{S}) &\approx& \frac{2E_0 \bar{S}}{E_0 + K_M + \bar{S} + \sqrt{(E_0+K_M+\bar{S})^2-4E_0 \bar{S}}}. \label{eqn: TQSSA Conditional} \nonumber\\
\end{eqnarray}

The uniform validity domain of the tQSSA is $k_{-1} \gg k_2$. However, any one of the following domains is also sufficient for the validiy of the tQSSA \cite{tzaf}:

\begin{eqnarray}
E_0 &\ll& K_M + S_0, \label{eqn: TQSSA domain1}\\
S_0 &\ll& E_0 + K_M   \text{ and } k_{-1} \gg k_2, \label{eqn: TQSSA domain2}\\
E_0 &\gg& S_0 \text{ and }  E_0 \gg K_M \text{ and } k_{-1} \ll k_2 \label{eqn: TQSSA domain3}.
\end{eqnarray}

In order to investigate the stochastic behavior of the MM scheme, a chemical master equation is introduced. This leads to a reduced master equation for substrate or total substrate molecules, providing an approximation to the exact solution.

In \cite{bar} Barik \textit{et al.} introduce a general method of dealing with the stochastic tQSSA and derive the following reduced master equation for the MM scheme:

\begin{eqnarray}
\frac{d P(\bar{s};t)}{dt} = k_2 \langle c|\bar{s}+1 \rangle P(\bar{s} + 1;t) - k_2 \langle c|\bar{s} \rangle P(\bar{s};t), \label{eqn: TQSSA Barik} \nonumber\\
\end{eqnarray}

where $\langle c|\bar{s} \rangle$ is the conditional expectation of the number of complex molecules at the quasi-steady-state. The authors then substitute $\langle c|\bar{s} \rangle$ with the quasi-steady-state value of the complex $c(\bar{s})$ from the tQSSA obtained by Tzafriri \cite{tzaf}. To support the validity of the method, they derive a recurrence relation for the quasi-steady-state probability distribution $P(c|\bar{s})$:

\begin{eqnarray}
P(c|\bar{s})= \prod_{i=1}^{c} {\frac{k_1}{k_{-1}} \frac{(\bar{s}-i+1)(e_0-i+1)}{i}} P(0|\bar{s}), \nonumber\\
0 \le c \le \text{min}(\bar{s},e_0) \label{e13}
\end{eqnarray}

and compare $\langle c|\bar{s}\rangle$ obtained from Equation (\ref{e13}) with the tQSSA result for a specific choice of parameters. Seeing an almost exact matching, the authors conclude that $\langle c|\bar{s} \rangle$ acquired from Equation (\ref{e13}) ubiquitously substitutes the tQSSA result given by Equation (\ref{eqn: TQSSA Conditional}). However, the mere fact that Equation (\ref{e13}) does not contain the Michaelis-Menten constant $K_M$ while Equation (\ref{eqn: TQSSA Conditional}) does, already suggests an inconsistency.

In what follows we provide a derivation of the stochastic tQSSA for the MM scheme, based on the underlying assumptions of the tQSSA. Unlike the argument presented by Barik \textit{et al.} \cite{bar} which holds for a limited choice of parameters, our approach holds for the entire tQSSA parameter domain.

\section{Stochastic tQSSA}

We begin by introducing the chemical master equation for the total substrate $\bar{s} = s+c$ and the complex $c$.

\begin{eqnarray}
\frac{dP(\bar{s},c; t)}{dt} = &-& \big[k_1 (\bar{s}-c)(e_0 - c)+(k_{-1}+k_2)c]P(\bar{s},c;t) \nonumber\\
&+& k_1 (\bar{s}-c+1)(e_0-c+1)P(\bar{s}, c-1; t) \nonumber\\
&+& k_{-1}(c+1)P(\bar{s},c+1; t) \nonumber\\
&+& k_2 (c+1) P(\bar{s}+1, c+1; t). \label{e23}
\end{eqnarray}

Summing Equation (\ref{e23}) over $c$, we obtain:

\begin{eqnarray}
\frac{dP(\bar{s};t)}{dt} = -k_2 \langle c| \bar{s} ; t\rangle P(\bar{s};t) + k_2 \langle c| \bar{s} + 1 ; t \rangle P(\bar{s}+1;t), \label{e24} \nonumber\\
\end{eqnarray}

where

\begin{eqnarray}
\langle c | \bar{s} ; t \rangle = \sum_c c P(c| \bar{s};t).
\end{eqnarray}

\begin{figure}
\centerline{\includegraphics[width=0.95\columnwidth]{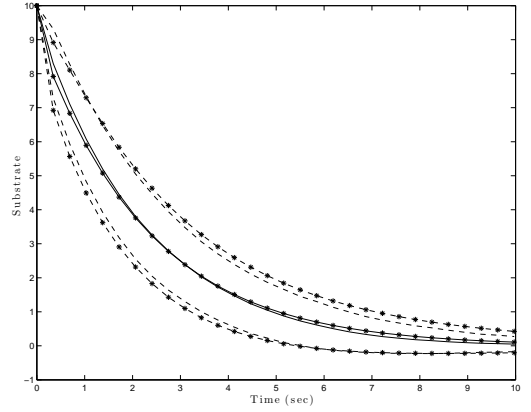}}
\caption{A comparison of the dynamics of the substrate molecules obtained from the Gillespie simulation of the scheme (\ref{MMK}) with that obtained from the stochastic tQSSA given by Equation (\ref{eqn: stochastic_tqssa}). The solid line represents the exact mean and the dashed lines represent one standard deviation away from the mean. The starred lines represent the tQSSA mean and $\pm 1$  standard deviation away from the mean. Parameters used: $k_1 = 0.1s^{-1}$, $k_{-1}=5 s^{-1}$, $k_2 = 5s^{-1}$, $E_0$=10, $S_0$=10. TQSSA condition (\ref{eqn: TQSSA domain1}) is met: $E_0 \ll S_0+ K_M$ }
\label{fig: tqssa vs. gill}
\end{figure}

The evolution of the complex $c$ for the given number of the total substrate $\bar{s}$ is governed by:

\begin{eqnarray}
\frac{dP(c|\bar{s};t)}{dt} = &-& \big[k_1 (\bar{s}-c)(e_0 - c)+(k_{-1}+k_2)c]P(c|\bar{s};t) \nonumber\\
&+& k_1 (\bar{s}-c+1)(e_0-c+1)P(c-1|\bar{s}; t) \nonumber\\
&+& (k_{-1}+k_2)(c+1)P(c+1|\bar{s}; t). \label{e16}
\end{eqnarray}

Multiplying both sides of Equation (\ref{e16}) by $c$ and summing over $c$, we obtain:

\begin{eqnarray}
\label{e17}
\frac{d \langle c | \bar{s} ;t \rangle }{dt} = 
 &\text{ }&  k_1 \big(\bar{s} e_0 - (e_0 + \bar{s}) \langle c|\bar{s};t \rangle   + \langle c | \bar{s} ; t \rangle^2 + \sigma^2_{c|\bar{s}} (t) \big) \nonumber\\
&-&  (k_{-1}+k_2) \langle c | \bar{s} ; t \rangle, \nonumber\\
\end{eqnarray}

where

\begin{eqnarray}
\sigma^2_{c|\bar{s}} (t)  = \sum_c (c - \langle c | \bar{s} ; t \rangle) ^2 P(c|\bar{s};t).
\end{eqnarray}

Applying the tQSSA, we set the right side of Equation (\ref{e17}) equal to zero, obtaining:

\begin{eqnarray}
\langle c| \bar{s} ; t\rangle^2  - r(\bar{s}) \langle c| \bar{s} ; t\rangle + e_0 \bar{s} \bigg( 1 + \frac{\sigma^2_{c|\bar{s}} (t)}  {e_0 \bar{s}} \bigg) = 0, \label{e29}
\end{eqnarray}

where 

\begin{equation}
r(\bar{s}) = K_M + e_0 + \bar{s}. \label{e30}
\end{equation}

\begin{figure}
\centerline{\includegraphics[width=0.95\columnwidth]{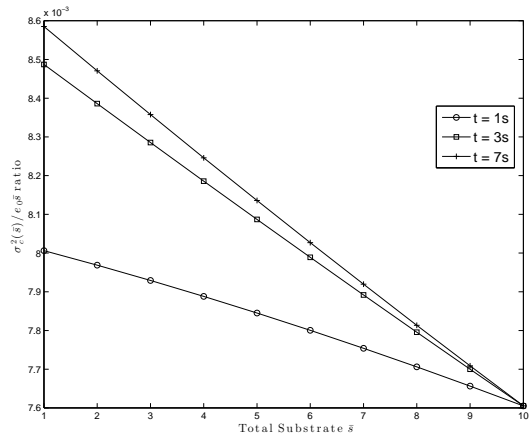}}
\caption{The ratio $\sigma^2 _{c|\bar{s}}/e_0 \bar{s}$ appearing in Equation (\ref{e29}) as a function of $\bar{s}$ at three different times, obtained from the exact solution of the original system (\ref{e23}). Parameters used: $k_1 = 0.1s^{-1}$, $k_{-1}=5s^{-1}$, $k_2 = 5s^{-1}$, $E_0$=10, $S_0$=10. TQSSA condition (\ref{eqn: TQSSA domain1}) is met: $E_0 \ll S_0+ K_M$ }
\label{fig: ratio}
\end{figure}

Typically $\sigma^2_{c|\bar{s}} (t)$ is on the same order of $\langle c| \bar{s} ;t\rangle$, which, in turn, does not exceed the initial number of enzyme molecules $e_0$. Thus, considering also the total substrate $\bar{s}$ term in the denominator, we can safely neglect the ratio $\sigma^2_{c|\bar{s} } (t)/e_0\bar{s}$ in Equation ({\ref{e29}}). (Fig. \ref{fig: ratio} depicts the ratio as a function of the total substrate $\bar{s}$ for three different times at the quasi-steady-state. As we can see, in all three cases it is much smaller than $1$ for all values of $\bar{s}$.) Solving for $\langle c | \bar{s}; t \rangle$ at the quasi-steady-state, we obtain:

\begin{eqnarray}
\langle c | \bar{s} \rangle &\approx& \frac{r(\bar{s}) - \sqrt{r^2(\bar{s})-4e_0 \bar{s}}}{2} \nonumber\\
&=& \frac{2e_0 \bar{s}}{e_0 + K_M + \bar{s} + \sqrt{(e_0 + K_M + \bar{s})^2 - 4e_0 \bar{s}}}. \label{e31} \nonumber\\
\end{eqnarray}

This expression matches exactly with the tQSSA result (\ref{eqn: TQSSA Conditional}) obtained by Tzafriri \cite{tzaf}, meaning that the substitution of the propensity function with the tQSSA rate function in Equation (\ref{e24}) is justified under the tQSSA conditions. This results in the reduced master equation for the total subtrate $\bar{s}$:

 \begin{eqnarray}
\frac{dP(\bar{s};t)}{dt} = -k_2 \langle c| \bar{s} \rangle P(\bar{s};t) + k_2 \langle c| \bar{s} + 1 \rangle P(\bar{s}+1;t), \label{eqn: stochastic_tqssa} \nonumber\\
\label{eqn: vahe tqssa}
\end{eqnarray}

where $\langle c | \bar{s} \rangle$ depends only on the total substrate $\bar{s}$ and is given by Equation (\ref{e31}). Fig. \ref{fig: tqssa vs. gill} compares the dynamics of the number and of the variance of substrate modecules obtained from the Gillespie simulation with those obtained from the stochastic tQSSA given by Equation (\ref{eqn: vahe tqssa}).

Applying the tQSSA ($\dot{P}(c|\bar{s};t) \approx 0$) on Equation (\ref{e16}), we can also derive a recurrence relation for the quasi-steady-state probability distribution $P(c|\bar{s})$:

\begin{eqnarray}
\label{e32}
P(c|\bar{s})= \prod_{i=1}^{c} {\frac{1}{K_M} \frac{(\bar{s}-i+1)(e_0-i+1)}{i}} P(0|\bar{s}), \nonumber\\
0 \le c \le \text{min}(\bar{s},e_0).
\end{eqnarray}

\begin{figure}
\centerline{\includegraphics[width=0.95\columnwidth]{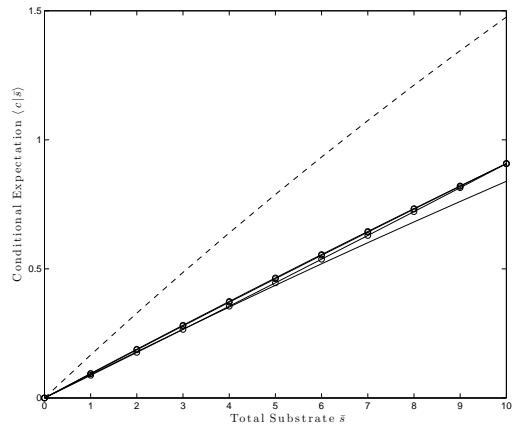}}
\caption{A comparison of the conditional expectation $\langle c | \bar{s} \rangle$ obtained from the recursive relation (\ref{e13}) and Equation (\ref{e31}) with that obtained from the exact solution of the original system (\ref{e23}) at three different quasi-steady-state moments: t = 1s, 3s, 7s. The dashed line represents Equation (\ref{e13}), the solid line - Equation (\ref{e31}), and the circles lines - the exact solution. Parameters used: $k_1 = 0.1s^{-1}$, $k_{-1}=5s^{-1}$, $k_2 = 5s^{-1}$, $E_0$=10, $S_0$=10. TQSSA condition (\ref{eqn: TQSSA domain1}) is met: $E_0 \ll S_0+ K_M$ }
\label{fig: barik vs vahe}
\end{figure}

Notice that the relation (\ref{e32}) is very similar to the relation (\ref{e13}), obtained by Barik \textit{et al.} \cite{bar}. The only difference is in the front coefficient, which in our case is $1/K_M$, whereas in Equation (\ref{e13}) it is $k_1/k_{-1}$. That is why the proof presented by Barik \textit{et al.} is valid only when $K_M \approx k_{-1}/k_1$, equivalent to $k_2/k_{-1} \ll 1$. Fig. \ref{fig: barik vs vahe} demonstrates that $\langle c| \bar{s} \rangle$ given by Equation (\ref{e31}) matches with $\langle c | \bar{s} ; t \rangle$ obtained from the exact solution at three different quasi-steady-state moments (t = 1s, 3s, 7s) much more closely than $\langle c | \bar{s} \rangle$ obtained by the recursive relation (\ref{e13}) suggested by Barik \textit{et al.} This means that the method of Barik \textit{et al.} is indeed limited to the case $k_{-1} \gg k_2$.

\section{Discussion}

In this paper we presented a derivation of the stochastic tQSSA for the MM scheme that was valid for the entire parameter range of the tQSSA. Substrate dynamics obtained from the reduced master equation was in perfect agreement with the Gillespie simulation of the MM scheme. We also showed that the method suggested by Barik \textit{et al.} \cite{bar} was limited to the special case $k_{-1} \ll k_2$. To our knowledge, our derivation of the stochastic tQSSA is a novel contribution. All code used in this manuscript is available as open source on \url{https://github.com/galstyan/stochasticTQSSA}.

We would also like to mention that an alternative approach of reducing the stochastic descriptions of biochemical networks was recently suggested by Thomas \textit{et al.} \cite{grima}. They used the method called slow-scale linear noise approximation (ssLNA) to evaluate the intrinstic noise present in enzyme networks. In the limit of large system size, the ssLNA appears to give better approximations to the stochastic MM scheme.

\section{Acknowlegements}

The present work benefited from valuable comments and assistance of Professor Chris Wiggins.

\renewcommand{\theequation}{A.\arabic{equation}}
\setcounter{equation}{0}

\end{document}